\documentclass[10pt,letterpaper]{article}
\usepackage{mathrsfs}
\usepackage{opex3}
\usepackage{amssymb}
\usepackage{amsmath}
\begin{document}

%%%%%%%%%%%%%%%%%% title page information %%%%%%%%%%%%%%%%%%
\title{Deterministic generation of multiparticle entanglement in a coupled cavity-fiber system}

\author{Peng-Bo Li* and Fu-Li Li}

\address{MOE Key Laboratory for Nonequilibrium Synthesis and Modulation of Condensed Matter,\\
Department of Applied Physics, Xi'an Jiaotong University, Xi'an
710049, China}

\email{lipengbo@mail.xjtu.edu.cn} %% email address is required



\begin{abstract}
We develop a one-step scheme for generating multiparticle entangled states between two cold atomic clouds in distant cavities coupled by an optical fiber. We show that, through suitably choosing the intensities and detunings of the fields and precisely tuning the time evolution of the system, multiparticle entanglement between the separated atomic clouds can be engineered deterministically, in which quantum manipulations are insensitive to the states of the cavity and losses of the fiber. The experimental feasibility of this scheme is analyzed based on recent experimental advances in the realization of strong coupling between cold $^{87}$Rb clouds and fiber-based cavity.
This scheme may open up promising perspectives for implementing quantum communication and networking with coupled cavities connected by optical fibers.
\end{abstract}

\ocis{(270.5585) Quantum information and processing; (270.5580)  Quantum electrodynamics } % REPLACE WITH CORRECT OCIS CODES FOR YOUR ARTICLE

%%%%%%%%%%%%%%%%%%%%%%% References %%%%%%%%%%%%%%%%%%%%%%%%%

%%%%%%%%%%%%%%%%%%%%%%%%%%  body  %%%%%%%%%%%%%%%%%%%%%%%%%%
\section{Introduction}
Multiparticle entangled states  are
indeed valuable resources, which not only can be employed to test quantum nonlocality against local hidden variable theory in fundamental physics \cite{Bell,GHZ}, but also
have practical applications in quantum
information processing \cite{QI}, such as quantum communication and computation. Typical such entangled states are GHZ
states \cite{GHZ}, W states \cite{W} and cluster states \cite{cluser}.
It is known that entangled states become
increasingly susceptible to environmental interactions
if the number of particles increases. Therefore, an important
practical challenge is the design of robust and most importantly
decoherence-resistant mechanisms for its generation. Generating multiparticle entangled states has been proposed or even experimentally demonstrated in different physical systems, such as atomic ensembles in free space \cite{RMP-82-1041}, trapped ions \cite{RMP-82-1209,nature-453-1008}, cold atoms in optical lattice \cite{AP-315}, and cavity QED \cite{QED1,QED2}. Among various excellent systems,
cavity QED \cite{QED1,QED2} offers one of the most promising and qualified candidates
for quantum state engineering and quantum information
processing \cite{prl-78-3221,prl-98-193601,prl-90-027903,pra-79-042339,pra-80-042319,josab-1,OE1}, particularly, for applications in quantum networking \cite{N-453-1023}, quantum communication, and distributed quantum computation, since atoms trapped in optical cavities are the natural candidates for quantum nodes, and these nodes can be connected by quantum channels such as optical fibers \cite{prl-78-3221,N-453-1023}. Quantum information is generated, processed
and stored locally in each node, which is connected by optical fibers, and is transferred between different nodes via photons through the fibers. Recently, with the development of experimental realization of strong coupling between cold atoms and fiber-based cavity \cite{nature-450-272,apl-87-211106,njp-12-065038-2010}, the coupled cavity-fiber system has been extensively investigated \cite{prl-79-5242-1997,prl-96-010503,pra-75-012324,pra-75-062320,pra-79-044304,arxiv-1009.1011,OE2}. Since entangled distant atomic clouds are the  building blocks for quantum network and quantum communication, it is desirable to develop a one-step scheme for generating multiparticle entangled states between two cold atomic clouds in distant cavities coupled by an optical fiber.

In this work, we describe a method to construct multiparticle entangled states of the form $\vert\psi_s\rangle=\frac{1}{\sqrt{2}}[e^{i\phi_0}\vert 000\cdots000\rangle_1\vert 111\cdots111\rangle_2+e^{i\phi_1}\vert 111\cdots111\rangle_1\vert 000\cdots000\rangle_2]$ or $\vert\psi_a\rangle=\frac{1}{\sqrt{2}}[e^{i\phi_0}\vert 000\cdots000\rangle_1\vert 000\cdots000\rangle_2+e^{i\phi_1}\vert 111\cdots111\rangle_1\vert 111\cdots111\rangle_2]$ in a coupled cavity-fiber system, where $\vert 000\cdots000\rangle_j$ and $\vert 111\cdots111\rangle_j$ ($N$ terms) are product states
describing $N$ atoms in the j\emph{th} cavity which are all in the (same) internal state
$\vert0\rangle$ or $\vert1\rangle$. This method can be implemented in one step and in a deterministic fashion. Through suitably choosing the intensities and detunings of the
fields and precisely controlling the dynamics of the system, the target entangled states can be engineered, which are immune to the spontaneous emission of the atoms and losses of the fiber, and independent of the states of the cavities. As an application, we also discuss how to use the produced entangled atomic state $\vert\psi_s\rangle$ to generate the so called NOON state of the cavity modes \cite{prl-99-053602}, i.e., $\vert N0 0N\rangle=(\vert N\rangle_1\vert 0\rangle_2+\vert 0\rangle_1\vert N\rangle_2)/\sqrt{2}$, where $\vert n\rangle_j(j=1,2)$ is Fock state for the respective cavity mode. This state is important
for quantum lithography and Heisenberg-limited
interferometry with photons. We should emphasize that the NOON state, which is mode-entangled, is different from the multiparticle atomic entangled states proposed in this scheme. The generated NOON state in itself is a two-mode entangled state, in which the entanglement is between two different cavity modes. However, the generated atomic entangled states are particle-entangled.
The experimental feasibility and technical demands of this scheme are analyzed based on recent experimental advances in the strong coupling between cold $^{87}$Rb clouds and fiber-based cavity.
Implementing the proposal in experiment would be an important step toward quantum
communication and networking with atomic clouds in distant cavities connected by optical fibers.

\section{Generation of the entangled states}
Consider two cold atomic clouds trapped in two distant optical cavities coupled by a short optical fiber--roughly a few hundred
of them trapped in each cavity--such that the
spatial extent of the cloud is much smaller than the wavelength
of the light---Lamb-Dicke regime. In this case the couplings between the atoms and the light fields are nearly homogeneous, which no longer depend on the individual coordinates
of the atoms but only on the collective pseudo-spin coordinate \cite{PR-408-315}. However, more realistic considerations will be given in the final discussions on the feasibility of this scheme. For simplicity, the atomic number in each cloud is assumed to be $N$, which however is not the necessary condition. The atomic level structure and the couplings for the atoms with cavity mode and external driving laser fields are shown in Fig.1.
\begin{figure}[htbp]
\centering
\includegraphics[bb=88 370 572 745,totalheight=2.3in,clip]{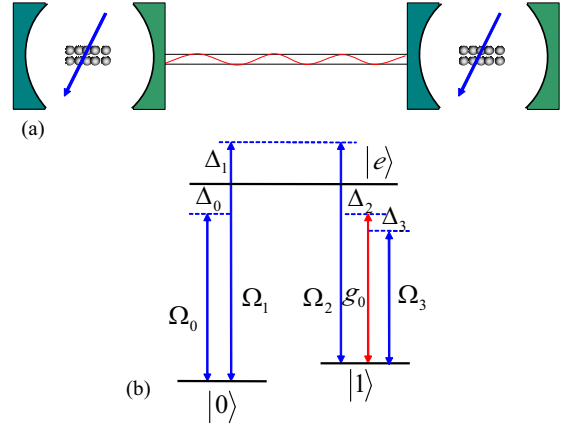}
\caption{(a) Experimental setup. Two distant cold atomic gases are
trapped in separate cavities connected by an optical fiber. The two cavity
modes are coupled to the fiber mode with the coupling strength $\nu$.
(b) Atomic level structure with
couplings to the cavity mode and driving laser fields.}
\end{figure}
Each atom has the level structure of three-level $\Lambda$ configuration, with two ground states $\vert0\rangle$ and $\vert1\rangle$, and an upper state $\vert e\rangle$. The atoms are driving by four laser fields and dispersively coupling to the cavity mode, establishing two distinct Raman transitions.
The cavity mode (frequency $\nu_0$) and the driving field of frequency $\omega_0$ couple the states $\vert0\rangle$ and $\vert1\rangle$ via the upper state $\vert e\rangle$ with the coupling constant $g_0$ and Rabi frequency $\Omega_0$ (complex), which is assumed to have a small detuning $\delta$ from the two-photon Raman resonance. The laser fields of frequencies
$\omega_1$ and $\omega_2$ act on the transitions $\vert0\rangle\leftrightarrow\vert e\rangle$ and $\vert1\rangle\leftrightarrow\vert e\rangle$ with Rabi frequencies $\Omega_1$ and $\Omega_2$ (complex), which satisfy the usual Raman resonance. To
cancel the Stark shift for level $\vert1\rangle$ caused by the classical laser field (frequency $\omega_2$), the transition  $\vert1\rangle\leftrightarrow\vert e\rangle$ is also driven by a third classical laser field of frequency $\omega_3$ with Rabi frequency $\Omega_3$ (complex).
The corresponding detunings for the related transitions are $\Delta_0=\omega_{e0}-\omega_0,\Delta_1=\omega_{e0}-\omega_1=\omega_{e1}-\omega_2,\Delta_2=\omega_{e1}-\nu_0=\Delta_0-\delta$, and $\Delta_3=\omega_{e1}-\omega_3$, where $\omega_{e0},\omega_{e1}$ are the transition frequencies for the atoms.
In the interaction
picture, the Hamiltonian describing the atom-field interaction
is (let $\hbar=1$)
\begin{eqnarray}
\label{H1}
\mathscr{H}_1
&=&\sum_{j=1,2}\{\sum_{n=1}^N[\Omega_0\vert e_n^j\rangle\langle 0_n^j\vert e^{i\Delta_0t}+\Omega_1\vert e_n^j\rangle\langle 0_n^j\vert e^{i\Delta_1t}\nonumber\\
&&+\Omega_2\vert e_n^j\rangle\langle 1_n^j\vert e^{i\Delta_1t}+\Omega_3\vert e_n^j\rangle\langle 1_n^j\vert e^{i\Delta_3t}
+
g_0\hat{a}_j\vert e_n^j\rangle\langle 1_n^j\vert e^{i(\Delta_0-\delta)t}]\}+\mbox{H.c.},\nonumber\\
\end{eqnarray}
where $\hat{a}_j$ is the annihilation operator for the j\emph{th} cavity mode. We proceed to derive the effective Hamiltonian of the atom-field interaction under the following conditions: (i) $|\Delta_0|,|\Delta_1|,|\Delta_2|,|\Delta_3|,|\Delta_0-\Delta_1|,|\Delta_1-\Delta_2|,|\Delta_2-\Delta_3|\gg|\Omega_0|,|\Omega_1|,|\Omega_2|,|g_0|,|\delta|$; (ii) $|\Omega_2|,|\Omega_3|\gg |g_0|;$ (iii) $ -|\Omega_0|^2/|\Delta_0|\sim|\Omega_1|^2/|\Delta_1|,-|\Omega_2|^2/|\Delta_1|\sim|\Omega_3|^2/|\Delta_3|$. Condition (i) guarantees that the dominate process is the two-photon Raman transitions between the states $\vert 0\rangle$ and $\vert 1\rangle$ via the excited state $\vert e\rangle$. Condition (ii) ensures that the terms
proportional to $|g_0|^2$ can be neglected. Condition (iii)
can completely cancel the energy shifts (Stark shifts) and the related
terms. In what follows we present the derivations in more detail. Using the time-averaged method \cite{James}, from the condition (i) we can adiabatically eliminate the excited state $\vert e\rangle$ and obtain
\begin{eqnarray}
\label{H1-1}
\mathscr{H}_1
&=&-\sum_{j=1,2}\{\sum_{n=1}^N[(\frac{|\Omega_0|^2}{\Delta_0}+\frac{|\Omega_1|^2}{\Delta_1})\vert 0_n^j\rangle\langle 0_n^j\vert +(\frac{|\Omega_2|^2}{\Delta_1}+\frac{|\Omega_3|^2}{\Delta_3}+\frac{|g_0|^2}{\Delta_2}\hat{a}_j^\dag \hat{a}_j)\vert 1_n^j\rangle\langle 1_n^j\vert\nonumber\\
&&+\frac{\Omega_1\Omega_2^*}{\Delta_1}\vert 1_n^j\rangle\langle 0_n^j\vert +\frac{\Omega_0g_0^*}{\Delta_0}\hat{a}_j^\dag\vert 1_n^j\rangle\langle 0_n^j\vert e^{i\delta t}]\}+\mbox{H.c.},
\end{eqnarray}
Then from conditions (ii) and (iii), the first two terms in equation (\ref{H1-1}) can be discarded.
Thus we obtain
\begin{eqnarray}
\label{H1-2}
\mathscr{H}_1
&=&\sum_{j=1,2}[\beta S^+_j+\beta^*S^-_j]+\sum_{j=1,2}[ \Lambda a_j^\dag e^{i\delta t}S^+_j+\Lambda^* a_je^{-i\delta t}S^-_j]
\end{eqnarray}
with $\beta=-\frac{\Omega_1\Omega_2^*}{\Delta_1}$, $\Lambda=-\frac{\Omega_0g_0^*}{\Delta_0}$, $S^+_j=\sum_{n=1}^N\vert 1_n^j\rangle\langle 0_n^j\vert$, and $S^-_j=(S^+_j)^\dag$.

We now consider the coupling of the cavity modes to the fiber modes. The number of longitudinal modes
of the fiber that significantly interact with the corresponding
cavity modes is on the order of $l\bar{\nu}/2\pi c$, where $l$ is the
length of the fiber and $\bar{\nu}$ is the decay rate of the cavity
fields into the continuum of the fiber modes. If we consider the short fiber limit $l\bar{\nu}/2\pi c\leq1$, then only one resonant mode $b$ of the
fiber interacts with the cavity modes \cite{prl-96-010503}. Therefore, in this case the interaction Hamiltonian describing the coupling between the cavity modes and the fiber mode is
\begin{eqnarray}
\label{H2}
\mathscr{H}_{c,f}&=&\nu b(\hat{a}_1^\dag+e^{i\varphi}\hat{a}_2^\dag)+\mbox{H.c.},
\end{eqnarray}
where $\nu$ is the cavity-fiber
coupling strength, and $\varphi$ is the phase due to propagation
of the field through the fiber. Define three normal bosonic modes $c,c_1,c_2$ by the canonical transformations
$c=\frac{1}{\sqrt{2}}(\hat{a}_1-e^{i\varphi}\hat{a}_2),c_1=\frac{1}{2}(\hat{a}_1+e^{i\varphi}\hat{a}_2+\sqrt{2}b),c_2=\frac{1}{2}(\hat{a}_1+e^{i\varphi}\hat{a}_2-\sqrt{2}b)$ \cite{prl-96-010503}. In terms of the bosonic modes $c,c_1$ and $c_2$, the interaction Hamiltonian $\mathscr{H}_{c,f}$ is diagonal. We rewrite this Hamiltonian as $\mathscr{H}_0=\sqrt{2}\nu c_1^\dag c_1-\sqrt{2}\nu c_2^\dag c_2$. So the whole Hamiltonian
in the interaction picture is $\mathscr{H}=\mathscr{H}_0+\mathscr{H}_1$.

Now let us perform the unitary transformation $e^{i\mathscr{H}_0t}$, which leads to
\begin{eqnarray}
\label{H}
\mathscr{H}'&=&e^{i\mathscr{H}_0t}\mathscr{H}e^{-i\mathscr{H}_0t}\nonumber\\
&=&\sum_{j=1,2}[\beta S^+_j+\beta^*S^-_j]+\frac{\Lambda^*}{2}[e^{-i\sqrt{2}\nu t}c_1+e^{i\sqrt{2}\nu t}c_2+\sqrt{2}c]S^-_1e^{-i\delta t}\nonumber\\
&&+\frac{\Lambda^*}{2}[e^{-i\sqrt{2}\nu t}c_1+e^{i\sqrt{2}\nu t}c_2-\sqrt{2}c]S^-_2e^{-i\delta t}+
\mbox{H.c.}\nonumber\\
\end{eqnarray}
Then we make another unitary transformation $e^{i\mathscr{V}_0t}$, with
$\mathscr{V}_0=\sum_{j=1,2}[\beta S^+_j+\beta^* S^-_j]$. In the new atomic basis $\vert \pm\rangle_n^j=(\vert0_n^j\rangle\pm\vert1_n^j\rangle)/\sqrt{2}$, and under the strong driving limit $|\beta|\gg|\delta|,|\Lambda|,|\nu|$, we can bring the effective Hamiltonian (\ref{H}) to a new form under the rotating-wave approximation \cite{prl-90-027903}
\begin{eqnarray}
\label{H3}
\mathscr{H}_{eff}&=&\{\frac{\Lambda^*}{2}[e^{-i\sqrt{2}\nu t}c_1+e^{i\sqrt{2}\nu t}c_2+\sqrt{2}c]e^{-i\delta t}+\frac{\Lambda}{2}[e^{i\sqrt{2}\nu t}c_1^\dag+e^{-i\sqrt{2}\nu t}c_2^\dag+\sqrt{2}c^\dag]e^{i\delta t}\}\nonumber\\
&&\times\sum_{n=1}^N\frac{1}{2}(\vert +\rangle_n^1\langle+\vert-\vert -\rangle_n^1\langle-\vert)
+\{\frac{\Lambda^*}{2}[e^{-i\sqrt{2}\nu t}c_1+e^{i\sqrt{2}\nu t}c_2-\sqrt{2}c]e^{-i\delta t}
\nonumber\\&&+\frac{\Lambda}{2}[e^{i\sqrt{2}\nu t}c_1^\dag+e^{-i\sqrt{2}\nu t}c_2^\dag-\sqrt{2}c^\dag]e^{i\delta t}\}\sum_{n=1}^N\frac{1}{2}(\vert +\rangle_n^2\langle+\vert-\vert -\rangle_n^2\langle-\vert).
\end{eqnarray}
If we further assume
$|\nu|\gg\{|\delta|,|\Lambda|\}$, we can take
the rotating-wave approximation and safely neglect the nonresonant
modes $c_1,c_2$. At present, we can obtain the effective Hamiltonian
\begin{eqnarray}
\label{H4}
\mathscr{H}_{eff}&=&[ \theta/2 c^\dag e^{i\delta t}+\theta^*/2 ce^{-i\delta t}][S^1_x-S^2_x]\nonumber\\
&=&\Theta/2[c^\dag e^{i\delta t+i\theta_0}+\mbox{H.c.}][S^1_x-S^2_x],
\end{eqnarray}
with $\theta=\frac{\sqrt{2}\Lambda}{2}=\Theta e^{i\theta_0}$, and $S^j_x=S^+_j+S^-_j$. It is worth emphasizing that, as the dominant interacting mode $c$
has no contribution from the fiber mode $b$, the system gets
in this instance insensitive to fiber losses. By
using the magnus formula, the evolution operator $\mathscr{U}(t)$ is
found as \cite{pra-62-022311,prl-94-100502}
\begin{eqnarray}
\label{U1}
\mathscr{U}(t)&=&e^{-i\gamma(t)[S^1_x-S_x^2]^2}e^{[\alpha(t)c^\dag-\alpha^*(t)c][S^1_x-S_x^2]},
\end{eqnarray}
where $\gamma(t)=-(\Theta^2/4\delta^2)(\delta t-\sin\delta t)$, and $\alpha(t)=(\Theta/2\delta)(1-e^{i\delta t})e^{i\theta_0}$. If the interaction time
$\tau$ satisfies $\delta\tau=2K\pi$, the evolution operator for the interaction Hamiltonian (\ref{H4})
can be expressed as
\begin{eqnarray}
\label{U2}
\mathscr{U}(\tau)&=&e^{-i\lambda\tau[S^1_x-S_x^2]^2},
\end{eqnarray}
where $\lambda=-\Theta^2/4\delta$. Note that as this operator has no contribution from the cavity modes, thus in this instance the system gets insensitive to the states of the cavity modes, which allows the cavity modes to be in a
thermal state.

After obtaining the time evolution operator (\ref{U2}), we now switch to the problem of generating multiatom entangled states in the coupled cavity-fiber system. Assume that the initial state of the atoms in the first cavity is $\vert 000\cdots000\rangle_1$, and in the second cavity is
$\vert 111\cdots111\rangle_2$. We now show how to prepare the multiatom maximally entangled state $\vert\psi_s\rangle=\frac{1}{\sqrt{2}}[e^{i\phi_0}\vert 000\cdots000\rangle_1\vert 111\cdots111\rangle_2+e^{i\phi_1}\vert 111\cdots111\rangle_1\vert 000\cdots000\rangle_2]$. In such a situation, it is convenient to apply the spin
representation of atomic states, where a collective state of
all atoms is represented by $\vert J,M\rangle$ which is an eigenstate of
the $S_z$ operator \cite{prl-82-1835}. Then the atomic states $\vert 000\cdots000\rangle$ and $\vert 111\cdots111\rangle$ can be expressed as
$\vert N/2,-N/2\rangle$ and $\vert N/2,N/2\rangle$. Alternatively, we can expand the initial states in terms of the eigenstates of $S_x$: $\vert N/2,-N/2\rangle=\sum c_M\vert N/2,M\rangle_x$ and $\vert N/2,N/2\rangle=\sum c_M(-1)^{N/2-M}\vert N/2,M\rangle_x$. Then we can express the initial state of the whole system as $\vert N/2,-N/2\rangle_1\vert N/2,N/2\rangle_2=\sum [c_{M1}c_{M2}(-1)^{N/2-M_2}\vert N/2,M_1\rangle_x\vert N/2,M_2\rangle_x]$. After applying the evolution operator $U(\tau)$, the final state will be
\begin{eqnarray}
\label{E1}
\vert\psi_f\rangle&=&\sum_{M_1,M_2=-N/2}^{N/2}c_{M1}c_{M2}(-1)^{N/2-M_2}e^{-i\lambda(M_1-M_2)^2\tau}\nonumber\\
&&\vert N/2,M_1\rangle_x\vert N/2,M_2\rangle_x.
\end{eqnarray}
If we choose $\lambda\tau=\pi/2$, then we have
\begin{eqnarray}
\label{E2}
\vert\psi_f\rangle&=&\frac{1}{\sqrt{2}}\sum_{M_1,M_2=-N/2}^{N/2}c_{M1}c_{M2}(-1)^{N/2-M_2}[e^{-i\pi/4}+e^{i\pi/4}(-1)^{M_1-M_2}]\nonumber\\
&&\vert N/2,M_1\rangle_x\vert N/2,M_2\rangle_x\nonumber\\
&=&\frac{1}{\sqrt{2}}[e^{-i\pi/4}\vert N/2,-N/2\rangle_1\vert N/2,N/2\rangle_2\nonumber\\
&&+e^{i\pi/4}\vert N/2,N/2\rangle_1\vert N/2,-N/2\rangle_2].
\end{eqnarray}
In the original atomic basis, the produced entangled state of Eq. (\ref{E2}) corresponds to $\vert\psi_s\rangle=\frac{1}{\sqrt{2}}[e^{-i\pi/4}\vert 000\cdots000\rangle_1\vert 111\cdots111\rangle_2+e^{i\pi/4}\vert 111\cdots111\rangle_1\vert 000\cdots000\rangle_2]$. If we prepare the initial states of both clouds in $\vert 000\cdots000\rangle$, then after applying the evolution operator $\mathscr{U}(\tau)$, the final state will be
\begin{eqnarray}
\label{E3}
\vert\psi_f\rangle&=&\sum_{M_1,M_2=-N/2}^{N/2}c_{M1}c_{M2}e^{-i\lambda(M_1-M_2)^2\tau}\nonumber\\
&&\vert N/2,M_1\rangle_x\vert N/2,M_2\rangle_x.
\end{eqnarray}
If we choose $\lambda\tau=\pi/2$, then we have
\begin{eqnarray}
\label{E4}
\vert\psi_f\rangle&=&\frac{1}{\sqrt{2}}\sum_{M_1,M_2=-N/2}^{N/2}c_{M1}c_{M2}[e^{-i\pi/4}+e^{i\pi/4}(-1)^{M_1-M_2}]\nonumber\\
&&\vert N/2,M_1\rangle_x\vert N/2,M_2\rangle_x\nonumber\\
&=&\frac{1}{\sqrt{2}}[e^{-i\pi/4}\vert N/2,-N/2\rangle_1\vert N/2,-N/2\rangle_2\nonumber\\
&&+e^{i\pi/4}\vert N/2,N/2\rangle_1\vert N/2,N/2\rangle_2].
\end{eqnarray}
Eq. (\ref{E4}) corresponds to $\vert\psi_a\rangle=\frac{1}{\sqrt{2}}[e^{-i\pi/4}\vert 000\cdots000\rangle_1\vert 000\cdots000\rangle_2+e^{i\pi/4}\vert 111\cdots111\rangle_1\vert 111\cdots111\rangle_2].$
Entangled states of Eq. (\ref{E2}) and (\ref{E4}) are maximally entangled, which would have very interesting applications in long-distance quantum communication with atomic clouds in optical cavities connected by optical fibers, and can be used to test quantum nonlocality in fundamental physics. It is noted that the above discussions do not rely on whether the atomic number $N$ is odd or even, which is quite different from
previous studies \cite{pra-62-022311,prl-94-100502}.

At this stage, we illustrate how to produce the entangled state of the cavity modes $\vert N0 0N\rangle$. It is known that these entangled states are important
for quantum lithography and Heisenberg-limited
interferometry with photons. Using the generated atomic entangled state $\vert\psi_s\rangle$, we wish to produce the state $\vert N0 0N\rangle$. To this aim, we employ the stimulated Raman transitions between the atomic ground states $\vert0\rangle$ and $\vert1\rangle$. After preparing the two atomic clouds in the target entangled state $\vert\psi_s\rangle$, we switch off the driving laser fields of frequencies $\omega_1,\omega_2$ and the couplings of the cavity modes to the fiber modes. Then we have two entangled atomic clouds trapped in two separated cavities, where the couplings of the atoms to the driving laser field and cavity modes are the $\Lambda$ type.
Starting from the state $\vert\psi_s\rangle\vert0\rangle_1\vert0\rangle_2=\frac{1}{\sqrt{2}}[e^{-i\pi/4}\vert 000\cdots000\rangle_1\vert 111\cdots111\rangle_2+e^{i\pi/4}\vert 111\cdots111\rangle_1\vert 000\cdots000\rangle_2]\otimes\vert0\rangle_1\vert0\rangle_2$, we are able to steer the evolved state towards the target state $\frac{1}{\sqrt{2}}[e^{-i\pi/4}\vert N\rangle_1\vert 0\rangle_2+e^{i\pi/4}\vert 0\rangle_1\vert N\rangle_2]\otimes\vert 111\cdots111\rangle_1\vert 111\cdots111\rangle_2$ through stimulated Raman transitions $\vert 000\cdots000\rangle\vert 0\rangle_j\longrightarrow\vert 111\cdots111\rangle\vert N\rangle_j.$

\section{Technical considerations}
In the discussions above, we have assumed that a cloud of cold
atoms can be trapped in an optical cavity
and prepared in the ground states, and the atom-field coupling
strengths are uniform through the atomic cloud. We now
analyze these assumptions are reasonable with the state-of-the-
art technology in experiment.
The
preparation of the initial atomic states can be accomplished
through the well-developed optical pumping and adiabatic
population transfer techniques.
For a cloud of cold $^{87}$Rb atoms cooled in the $\vert F=1\rangle $ ground state and trapped inside an optical cavity,  this cloud can be prepared in the $\vert F=2\rangle $ ground state employing either the optical pumping or adiabatic
population transfer techniques \cite{RMP-70-1003}.
We note that recent experimental advances are achieved with a BEC or cold cloud of $^{87}$Rb atoms positioned deterministically anywhere within the
cavity and localized entirely within a single antinode of the standing-wave cavity field \cite{nature-450-272}. In the experiment each atom is identically and strongly coupled to the cavity mode, and a controlled tunable coupling rate has been realized. For a certain lattice
site, a well-defined and maximal atom-field coupling
could be achieved. Define the average atom-field coupling strength $\bar{g}=\sqrt{\int \frac{\rho(\textbf{r})}{N}|g(\textbf{r})|^2d\textbf{r}}$, where $\rho(\textbf{r})$ is the atomic density distribution, $g(\textbf{r})=g_0\cos(k_cz)\exp[-r_\perp^2/w^2]$ is the position-dependent single-atom
coupling strength (here $z$ and $r_\perp$
are respectively the longitudinal and transverse atomic coordinates,
and $w$ and $k_c$ are respectively the mode radius and wave
vector). For a Gaussian
cloud centered on a single lattice site with $N < 10^3$, in which the distribution can be considered point-like,
an average atom-field coupling strength $\bar{g}/2\pi\simeq200$ MHz can be obtained \cite{nature-450-272}. The homogeneous interaction condition requires that the
variation of the coupling strength in a single lattice must be very small, i.e., $\delta g/\bar{g}\ll1$, or $k_c\delta z\ll1$ and $\delta r_\perp\ll w$. Therefore, under these conditions all the atoms in the cavity have
the nearly same coupling strength and then collectively interact
with the cavity field. On the other hand, to avoid the direct interaction between
the atoms being in the ground state, the mean atom-atom distance
should be larger than the radius of the atom in the ground
state. For a combined trap with the shape of flat disk employed in the experiment, the mean atom-atom distance may be estimated as $d=\sqrt{\pi(\delta r_\perp)^2/N}$.
Consider an alkali-metal
atom with the valence electron in $ns$ state. The orbit
radius of the valence electron is approximately as
$r_g\sim n^2a_0$, where $a_0$ is the Bohr radius. This imposes a condition on the atomic number, i.e., $N<\pi(\delta r_\perp)^2/n^4a_0^2$. For $n=5,\delta r_\perp\simeq2\mu$m, when $N<10^3$, we have $d\gg r_g$, which implies the single-particle approximation or no-direct interaction
condition is valid.

We consider the traveling-plane-wave driving laser fields with electric fields $\vec{E}_i=\mathscr{\vec{E}}_ie^{i[\vec{k}_i\cdot \vec{r}-\omega_it]}(i=0,1,2,3)$. Then the actual coupling coefficients between the atoms and the laser fields have the spatial phases $e^{i\vec{k}_i\cdot \vec{r}_n}$ for the individual atom with the position coordinate $\vec{r}_n$. Therefore, the effective Hamiltonian (\ref{H1-1})
should be in a more general form as
\begin{eqnarray}
\label{H1-3}
\mathscr{H}_1
&=&-\sum_{j=1,2}\{\sum_{n=1}^N[\frac{\Omega_1\Omega_2^*}{\Delta_1}e^{i(\vec{k}_1-\vec{k}_2)\cdot \vec{r}^j_n}\vert 1_n^j\rangle\langle 0_n^j\vert +\frac{\Omega_0g_0^*}{\Delta_0}e^{i\vec{k}_3\cdot \vec{r}^j_n}\hat{a}_j^\dag\vert 1_n^j\rangle\langle 0_n^j\vert e^{i\delta t}]\}+\mbox{H.c.}
\end{eqnarray}
After performing a time-independent unitary transformation $U=\exp\{\frac{1}{2}{\sum_{j=1,2}\sum_{n=1}^N[i(\vec{k}_1-\vec{k}_2)\cdot \vec{r}^j_n(\vert 1_n^j\rangle\langle 1_n^j\vert-\vert 0_n^j\rangle\langle 0_n^j)]}\}$ \cite{pra-77-062327}, we can bring the Hamiltonian (\ref{H1-3}) to the form
\begin{eqnarray}
\label{H1-4}
\mathscr{H}_1
&=&-\sum_{j=1,2}\{\sum_{n=1}^N[\frac{\Omega_1\Omega_2^*}{\Delta_1}\vert 1_n^j\rangle\langle 0_n^j\vert +\frac{\Omega_0g_0^*}{\Delta_0}e^{i(\vec{k}_3-\vec{k}_1+\vec{k}_2)\cdot \vec{r}^j_n}\hat{a}_j^\dag\vert 1_n^j\rangle\langle 0_n^j\vert e^{i\delta t}]\}+\mbox{H.c.}
\end{eqnarray}
If the relative orientation of the three driving laser fields are adjusted in such a way that $\vec{k}_3-\vec{k}_1+\vec{k}_2=0$, then the spatial phase terms will not appear in the effective Hamiltonian. Thus our discussion in the above section still holds. The only difference is that the state vector for each atom should be redefined as $\vert 0_n\rangle\rightarrow e^{-i/2(\vec{k}_1-\vec{k}_2)\cdot \vec{r}_n}\vert0_n\rangle$ and $\vert 1_n\rangle\rightarrow e^{i/2(\vec{k}_1-\vec{k}_2)\cdot \vec{r}_n}\vert1_n\rangle$ \cite{pra-77-062327}. On the other hand, if the atoms in the trap move around a mean position, i.e., $\vec{r}_n\approx \bar{R}+\delta r_n$, and the deviation from the mean position is much smaller than the wavelength of the laser---Lamb-Dicke regime, i.e., $\delta r_n\ll \lambda_L$, then the spatial phase terms are global and can be absorbed into the complex Rabi frequencies.

\begin{figure} [htbp]
\centering
\includegraphics[bb=64 366 567 818,totalheight=3in,clip]{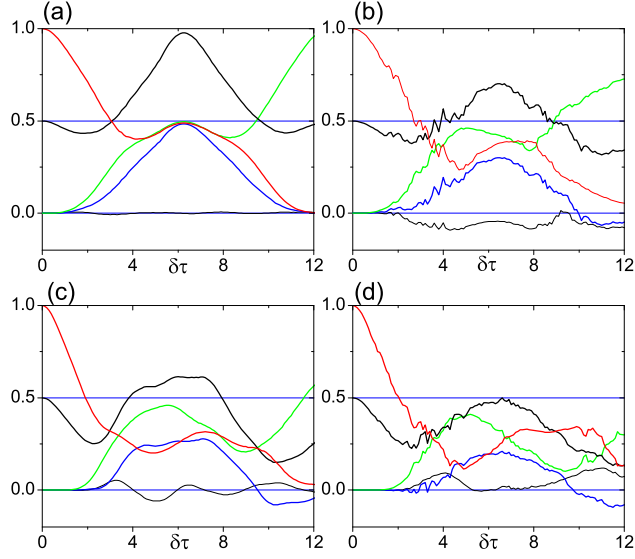}
\caption{Time evolution of the population and coherences of joint atomic ground states $\vert 000\cdots000\rangle$ and $\vert 111\cdots111\rangle$, as well as the fidelity. The first full curve (counted
from above at $\delta\tau<2$) is the population of the joint ground
state $\vert 000\cdots000\rangle$, the second one is the fidelity, the third one is population of the joint ground
state $\vert 111\cdots111\rangle$, the last two curves are the imaginary and real part of the off-diagonal elements of the atomic density matrix, respectively.
Results are displayed for different atomic numbers and cavity decay rates: (a) $N=2,\kappa_c=0.1g_0$; (b) $N=2,\kappa_c=0.5g_0$;
(c) $N=5,\kappa_c=0.1g_0$; (d) $N=5,\kappa_c=0.5g_0$. }
\end{figure}

Now let us verify the model and study the performance of this protocol under realistic
circumstances through numerical simulations. In a realistic experiment, the effect of the spontaneous
emission of the atoms and cavity and fiber losses on the produced entangled atomic states should be taken into account.
 We choose the parameters as $\Delta_0=100|g_0|,\Delta_1=-100|g_0|,|\Omega_0|=|g_0|,|\Omega_1|=10|g_0|,|\Omega_2|=10|g_0|,\nu=0.1|g_0|,\delta=0.01|g_0|$. The probability for the Raman transition $\vert 0\rangle\leftrightarrow\vert1\rangle$ induced by the classical field and the normal modes $c_1,c_2$ is on the order of $\mathscr{P}_f\sim|\Omega_0g_0|^2/(\Delta_0\nu)^2$. For fiber loss at a rate $\kappa_f$, we get the effective loss rate  $\Gamma_f=\kappa_f|\Omega_0g_0|^2/(\Delta_0\nu)^2\sim0.01\kappa_f$.
The occupation of the excited
state $\vert e\rangle$ can be estimated to be $\mathscr{P}_e\sim|\Omega_1|^2/|\Delta_1|^2$. Spontaneous
emission from the excited state at a rate $\gamma_e$ thus leads
to the effective decay rate $\Gamma_e=\gamma_e|\Omega_1|^2/|\Delta_1|^2\sim0.01\gamma_e$.
For this proposal, the probability for cavity excitation can be estimated as $\mathscr{P}_c\sim|\theta|^2/4\delta^2\simeq|\Omega_0g_0|^2/(8\delta^2\Delta_0^2)$. For cavity decay of photons at a rate
$\kappa_c$, this leads to an effective decay rate $\Gamma_c=\kappa_c|\Omega_0g_0|^2/(8\delta^2\Delta_0^2)\sim0.1\kappa_c$.
Therefore, The main decoherence effects in our scheme is due to cavity decay. Coherent interaction thus requires the preparing time $\tau\leq\{\Gamma_c^{-1},\Gamma_e^{-1},\Gamma_f^{-1}\}$.
In the following, we present
some numerical results to show how the cavity decay
affects the performance of this scheme. The evolution of
the system is governed by the following master equation
\begin{eqnarray}
\label{M}
\frac{d\rho}{dt}&=&-i[\mathscr{H},\rho]+\kappa_c\sum_{j=1}^2\mathscr{L}[\hat{a}_j]\rho
\end{eqnarray}
where the superoperator $\mathscr{L}[\hat{o}]=2\hat{o}\rho\hat{o}^\dag-\hat{o}^\dag\hat{o}\rho-\rho\hat{o}^\dag\hat{o}$.
To solve the
master equation numerically, we have used the Monte Carlo wave function formalism from the quantum trajectory method \cite{CPC-102-210}. All the simulations are performed under one trajectory with the atomic number amounting to ten. (To perform Monte Carlo simulations for the case of much larger number of atoms is time-consuming, which is even unable to be accomplished under present calculating conditions.)
In Fig. 2 we display the time evolution of the population and coherences of joint atomic ground states $\vert 000\cdots000\rangle$ and $\vert 111\cdots111\rangle$, as well as the fidelity $F=\langle\psi_f\vert\varrho_a\vert\psi_f\rangle$, where $\varrho_a$ is the final reduced density matrix of the atoms,  under different values for the parameter $\kappa_c$.  The atomic system starts from the joint ground state $\vert 000\cdots000\rangle$ in each cavity.
Fig. 2(a) and (b) show the calculated populations, coherences, and fidelity for the case of two atoms trapped in each cavity. We can see that (Fig. 2(a))
at the time $\tau=2\pi/\delta$ the state (\ref{E4}) is obtained with a fidelity higher than $99\%$ in the relatively strong coupling regime. However, from
Fig. 2(c) and (d), we find that, for the larger number of atoms even the same cavity decay rate leads to more pronounced degradation of the generated state. It seems that for large number of atoms the produced entangled states are much easier to be spoiled by losses, i.e., off-diagonal elements of the atomic density matrix may decay more rapidly. Therefore, for a cloud containing a few hundred
of atoms, to generate the target entangled states with a high fidelity may demand more stringent conditions.

For experimental implementation of this scheme with the fiber-connected coupled cavity system, the recent experimental setup of integrated fiber-based cavity system \cite{nature-450-272,apl-87-211106,njp-12-065038-2010} is particularly suitable. The core of this kind of cavity design is a concave, ultralow-roughness mirror surface fabricated directly on the end face of an optical fiber, or  two closely spaced fiber tips
placed face to face. Light couples directly in and out of the resonator through the optical fiber, which can be either single mode or multi-mode. Atoms or other emitters can be transported  into the cavity using optical lattices or other traps. For realization of this protocol, the best candidate for atomic system is $^{87}$Rb, with the ground states $\vert 0\rangle$ and $\vert 1\rangle$ corresponding to the $\vert5S_{1/2}\rangle$ hyperfine levels, and the excited state $\vert e\rangle$ corresponding to the $\vert5P_{1/2}\rangle$ sub-states. It is noted that the strong coupling of cold $^{87}$Rb clouds with the fiber-based cavity has been realized in recent experiments \cite{nature-450-272,apl-87-211106,njp-12-065038-2010}. The coupling strength between cavity mode and cold atoms ranges from $|g_0|/2\pi=10.6$MHZ to $|g_0|/2\pi=215$MHz.
We choose the cavity QED parameters as those in  Ref.\cite{nature-450-272}, $|g_0|/2\pi=215$MHz, $\kappa/2\pi=53$MHz, $\gamma/2\pi=3$MHz. Other experimental parameters can be chosen as $\Delta_0/2\pi=20\mbox{GHz},\Delta_1/2\pi=-20 \mbox{GHz},\Delta_3/2\pi=40 \mbox{GHz},|\Omega_0|/2\pi=200\mbox{MHz},|\Omega_1|/2\pi=2\mbox{GHz},|\Omega_2|/2\pi=2\mbox{GHz},|\Omega_3|/2\pi=3\mbox{GHz},\nu/2\pi=20\mbox{MHz},\delta/2\pi=2\mbox{MHz}$. The fiber length can be chosen as $L\lesssim1$m in most realistic
experimental situations. With the chosen parameters, for a few hundred
of cold atoms trapped in the cavity, the  time to prepare the entangled state $\tau\sim 0.5\mu$s.

\section{Conclusions}
To conclude, we have introduced an efficient scheme for producing multiparticle entangled states between distant atomic clouds in two cavities coupled by an optical
fiber. This proposal can be implemented in just one step, and is robust against the atomic spontaneous
emission and fiber losses. We have discussed the experimental feasibility of this proposal based on recent experimental advances in the strong coupling between cold $^{87}$Rb clouds and fiber-based cavity. We have also discussed how to use the generated multiatom states to produce the NOON
state for the two cavities. Experimental implementation of this proposal may offer a promising platform for implementing long-distance quantum communications with atomic clouds trapped in separated cavities connected by optical fibers.

\section*{Acknowledgments}
This work is supported by the National
Key Project of Basic Research Development under
Grant No.2010CB923102, and National Nature Science
Foundation of China under Grant No. 60778021. P.-B. L. acknowledges the support
from the New Staff Research Support Plan of Xi'an Jiaotong
University under No.08141015 and the useful discussions
with H.-Y. Li.


\begin{thebibliography}{99}

\bibitem{Bell} J. S. Bell, ``On the einstein-podolsky-rosen paradox,'' Physics \textbf{1}, 195-200 (1964).

\bibitem{GHZ} D. M. Greenberger, M.A. Horne, A. Shimony, and A. Zeilinger, ``Bell's theorem without inequalities,'' Am. J. Phys. \textbf{58},  1131-1143  (1990).

\bibitem{QI}M. A. Nielsen and I. L. Chuang, ``Quantum Computation and
Quantum Information,'' Cambridge University Press (Cambridge) (2000).

\bibitem{W} W. Dur, G. Vidal, and J. I. Cirac, ``Three qubits can be entangled in two inequivalent ways,'' Phys. Rev. A \textbf{62}, 062314 (2000).
\bibitem{cluser} H. J. Briegel and R. Raussendorf, ``Persistent entanglement in arrays of interacting particles,'' Phys. Rev. Lett. \textbf{86}, 910-913 (2001).

\bibitem{RMP-82-1041} For a review see, K. Hammerer, A. S. Sorensen,  and E. S. Polzik, ``Quantum interface between light and atomic ensembles,'' Rev. Mod. Phys. \textbf{82}, 1041-1093 (2010), and references therein.


\bibitem{RMP-82-1209} For a review see, L.-M. Duan and  C. Monroe, ``Colloquium: Quantum networks with trapped ions,'' Rev. Mod. Phys. \textbf{82}, 1209-1224 (2010), and references therein.

\bibitem{nature-453-1008}  R. Blatt and D. Wineland,  ``Entangled states of trapped atomic ions,'' Nature (London) \textbf{453}, 1008-1015 (2008).

\bibitem{AP-315} For a review see, D. Jaksch  and P. Zoller, ``The cold atom Hubbard toolbox,'' Ann. Phys.  \textbf{315}, 52-79 (2005), and references therein.

\bibitem{QED1} H. J. Kimble, ``Strong interactions of single atoms and photons in cavity QED,'' Phys. Scr. \textbf{T76}, 127-137 (1998).

\bibitem{QED2} H. Mabuchi and A. C. Doherty, ``Cavity quantum electrodynamics: coherence in Context,'' Science \textbf{298}, 1372-1377  (2002).


\bibitem{prl-78-3221} J. I. Cirac, P. Zoller, H. J. Kimble, and H. Mabuchi, ``Quantum state transfer and entanglement distribution among distant nodes in a quantum network,'' Phys. Rev. Lett. \textbf{78}, 3221-3224  (1997).

\bibitem{prl-98-193601} A. D. Boozer, A. Boca, R. Miller, T. E. Northup, and H. J. Kimble,  ``Reversible state transfer between light and a single trapped atom,'' Phys. Rev. Lett. \textbf{98}, 193601 (2007).

\bibitem{prl-90-027903} E. Solano, G. S. Agarwal, and H. Walther, ``Strong-driving-assisted multipartite entanglement in cavity QED,'' Phys. Rev. Lett. \textbf{90}, 027903 (2003).

\bibitem{pra-79-042339} P.-B. Li, Y. Gu, Q.-H. Gong, and G.-C. Guo, ``Quantum-information transfer in a coupled resonator waveguide,'' Phys. Rev. A \textbf{79}, 042339 (2009).

\bibitem{pra-80-042319} F. Mei, M. Feng,  Y.-F. Yu, and Z.-M. Zhang, ``Scalable quantum information processing with atomic ensembles and flying photons,'' Phys. Rev. A \textbf{80}, 042319 (2009).

\bibitem{josab-1} P.-B. Li, Y. Gu, Q.-H. Gong, and G.-C. Guo, ``Generation of two-mode entanglement between
separated cavities,''  J. Opt. Soc. Am. B \textbf{26}, 189-193 (2009).

\bibitem{OE1} S. Kang, Y. Choi, S. Lim, W. Kim, J.-R. Kim, J.-H. Lee, and K. An, ``Continuous control of the coupling constant in an atom-cavity system by using elliptic polarization and magnetic sublevels,'' Opt. Express. \textbf{18}, 9286-9302 (2010).


\bibitem{N-453-1023} H. J. Kimble, ``The quantum internet,'' Nature (London) \textbf{453}, 1023-1030 (2008).

\bibitem{nature-450-272} Y. Colombe, T. Steinmetz, G. Dubois, F. Linke, D. Hunger, and  J. Reichel, ``Strong atom-field coupling for Bose-Einstein
condensates in an optical cavity on a chip,'' Nature (London) \textbf{450}, 272-276 (2007).

\bibitem{apl-87-211106} M. Trupke, E. A. Hinds, S. Eriksson, E. A. Curtis, Z. Moktadir, E. Kukharenka, and M. Kraft, ``Microfabricated high-finesse optical cavity with open access and small volume,'' Appl. Phys. Lett. \textbf{87}, 211106 (2005).


\bibitem{njp-12-065038-2010} D. Hunger, T. Steinmetz, Y. Colombe, C. Deutsch, T. W. Hansch, and J. Reichel, ``A fiber Fabry-Perot cavity with high finesse,'' New J. Phys. \textbf{12} 065038 (2010).



\bibitem{prl-79-5242-1997}  T. Pellizzari, ``quantum networking with optical fibres,'' Phys. Rev. Lett. \textbf{79}, 5242-5245 (1997).

\bibitem{prl-96-010503} A. Serafini, S. Mancini, and S. Bose, ``Distributed quantum computation via optical fibers,'' Phys. Rev. Lett. \textbf{96}, 010503 (2006).
\bibitem{pra-75-012324} Z. Q. Yin and F. L. Li, ``Multiatom and resonant interaction scheme for quantum state transfer and logical gates between two remote cavities via an optical fiber,'' Phys. Rev. A \textbf{75}, 012324 (2007).

\bibitem{pra-75-062320} P. Peng and F. L. Li, ``Entangling two atoms in spatially separated cavities through both photon emission and absorption processes,'' Phys. Rev. A \textbf{75}, 062320 (2007).

\bibitem{pra-79-044304} Y. L. Zhou, Y. M. Wang, L. M. Liang, and C. Z. Li, ``Quantum state transfer between distant nodes of a quantum network via adiabatic passage,'' Phys. Rev. A \textbf{79},044304 (2009).

\bibitem{arxiv-1009.1011} J. Busch, and A. Beige, ``Generating single-mode behavior in fiber-coupled optical cavities,'' arXiv:1009.1011v2  (2010).

\bibitem{OE2} X.-Y. Lu, P.-J. Song,  J.-B. Liu, and  X. Yang, ``N-qubit W state of spatially separated single molecule magnets,'' Opt. Express. \textbf{17}, 14298-14311 (2010).


\bibitem{prl-99-053602} K. T. Kapale and J. P. Dowling,  ``Bootstrapping approach for generating maximally path-entangled photon states,'' Phys. Rev. Lett. \textbf{99}, 053602  (2007).

\bibitem{PR-408-315} T. Brandes, ``Coherent and collective quantum optical effects in mesoscopic
systems,'' Phys. Rep. \textbf{408},  315-474 (2005).

\bibitem{James}   D. F. V. James, ``Quantum computation with hot and cold ions: an assessment of proposed schemes,'' Fortschr. Phys. \textbf{48}, 823-837 (2000).

\bibitem{pra-62-022311} A. S{\o}rensen and K. M{\o}lmer, ``Entanglement and quantum computation with ions in thermal motion,'' Phys. Rev. A \textbf{62}, 022311 (2000).

\bibitem{prl-94-100502} S. L. Zhu, Z. D. Wang, and P. Zanardi, ``Geometric quantum computation and multiqubit entanglement with superconducting qubits inside a cavity,'' Phys. Rev. Lett. \textbf{94},  100502 (2005).

\bibitem{prl-82-1835} K. M{\o}lmer and A. S{\o}rensen, ``Multiparticle entanglement of hot trapped ions,''
 Phys. Rev. Lett. \textbf{82}, 1835-1838 (1999).

\bibitem{RMP-70-1003}  K. Bergmann, H. Theuer, and B. W. Shore, ``Coherent population transfer among quantum states
of atoms and molecules,''  Rev. Mod. Phys. \textbf{70}, 1003-1025 (1997).


\bibitem{pra-77-062327} I. E. Linington and N. V. Vitanov, ``Decoherence-free preparation of Dicke states of trapped ions by collective stimulated Raman adiabatic passage,'' Phys. Rev. A \textbf{77}, 062327 (2008).


\bibitem{CPC-102-210}  R. Schack and T. A. Brun, ``A C++ library using quantum trajectories to solve quantum
master equations,'' Comput. Phys. Commun. \textbf{102}, 210-228 (1997).




\end{thebibliography}
\end{document}